\documentstyle[preprint,aps,epsfig,epsf,rotate]{revtex}

\begin{document}
\tighten

\preprint{
\vbox{
\hbox{Presented~by~A.~W.~Thomas~at~the}
\hbox{Int.~Symposium~on~Nuclear~Physics}
\hbox{(Mumbai, Dec. 2000)}
\hbox{ADP-01-05/T440}
}}

\def\bge{\begin{equation}}
\def\ene{\end{equation}}
\newcommand{\nc}{\newcommand}
\nc{\non}{\nonumber}
\def\be{\begin{equation}}
\def\ee{\end{equation}}
\def\bga{\begin{eqnarray}}
\def\ena{\end{eqnarray}}
\def\eea{\end{eqnarray}}
\def\bg{\begin{eqnarray}}
\def\en{\end{eqnarray}}
\def\nn{\nonumber}
\def\ra{\rightarrow}
\def\la{\leftarrow}
\def\ua{\uparrow}
\def\da{\downarrow}
\def\del{\partial}
\def\qqbar{q \bar{q}}
\def\qbar{{\bar{q}}}
\def\ubar{{\bar{u}}}
\title{A New Slant on Hadron Structure} 
\author{W.~Detmold$^1$, D.~B.~Leinweber$^1$, W.~Melnitchouk$^{1,2}$, 
A.~W.~Thomas$^1$ and S.~V.~Wright$^1$}
\address{$^1$ Special Research Centre for the Subatomic Structure of
Matter, and Department of Physics and Mathematical Physics,
Adelaide University, 5005, Australia}
\address{$^2$ Jefferson Lab, 12000 Jefferson Avenue,
Newport News, VA 23606, USA}
\maketitle
\begin{abstract}
Rather than regarding the restriction of current lattice QCD simulations
to quark masses that are 5--10 times larger than those observed, we
note that this presents a wonderful opportunity to deepen our
understanding of QCD. Just as it has been possible to learn a great deal
about QCD by treating $N_c$ as a variable, so the study of hadron
properties as a function of quark mass is leading us to a much deeper
appreciation of hadron structure. As examples we cite recent progress in
using the chiral properties of QCD to connect hadron masses, magnetic
moments, charge radii and structure functions calculated at large quark
masses within lattice QCD with the values observed physically.
\end{abstract}
  
\section{Introduction}
In striving to understand the properties of QCD the generalization to
an arbitrary number of colours, $N_c$, particularly the limit $N_c
\rightarrow \infty$ (or ``large $N_c$'') has been extremely 
valuable. It has even proven possible to distinguish between models of
hadron structure and to guide the further developments of such models on
the basis of their large $N_c$ behaviour \cite{Cohen}. 
Until recently it has generally been regarded as an unfortunate
liability that current limitations on computer power restrict lattice
QCD simulations with dynamical fermions to large quark masses.
We would like to present a rather different
view concerning the lattice data at large quark masses. In particular,
we argue that like
the behaviour as a function of $N_c$, lattice results
as a function of quark mass offer extremely valuable new insights into
the nature of QCD and especially into hadron structure.

To be a little more quantitative, the restriction to large quark masses
in lattice simulations means typically 50 MeV or higher. 
Thus, in order to compare hadron
properties calculated on the lattice one has to extrapolate as a
function of quark mass (on top of all the other extrapolations, lattice
spacing, lattice size, etc.) all the way to the physical light quark
masses, around 5 or 6 MeV. 
Such extrapolations are complicated enormously by the fact that  
chiral symmetry is spontaneously broken in QCD.
The mass of the pion, which is the Goldstone boson corresponding to this
broken symmetry \cite{Pagels}, behaves as:
\be
m_\pi^2 \propto \bar{m}, \hspace{2cm} ({\rm with}\hspace{0.5cm}
\bar{m}=m_u=m_d\neq0) ,
\label{eq:1}
\ee
as the quark mass, $\bar{m}$, moves away from zero -- 
this is the Gell Mann-Oakes-Renner (GOR) relation.
While Eq.(1) is, in principle, only guaranteed for quark masses,
near zero, explicit lattice calculations show that it holds over an
enormous range, as high as $m_\pi \sim 1$GeV. For convenience, 
rather than measuring the deviation
from exact chiral symmetry using $\bar{m}$, which is scale dependent, we
shall use $m_\pi^2$.

In terms of $m_\pi$, current lattice 
calculations are typically restricted to pion masses
larger than 500 MeV, with some pioneering work reporting preliminary
results as low as 310 MeV. In order to compare these results with
experimental data on hadron properties it is necessary to extrapolate
the calculations at large pion masses to the physical value. In doing so
it is crucial to respect the constraints imposed by chiral symmetry in
QCD. In particular, as we discuss below, the existence of Goldstone
bosons necessarily leads to behaviour 
which is {\bf non-analytic} in the quark mass. 

The structure of this article is that we first explain the origin of the
non-analyticity associated with Goldstone boson loops.
We then explain, using the specific case of the nucleon mass, how this 
non-analytic structure has been incorporated into a new method for 
extrapolating hadron masses from the large values characteristic of
lattice calculations to the physical region.
The consequences of this for the sigma commutator are
then explained. Next we turn to recent results for baryon
electromagnetic properties. Finally we discuss the most recent
investigations of the proton structure function, especially the
importance of chiral symmetry in connecting existing calculations of
lattice moments with data. We conclude with a summary of the promised
insights into the nature of hadron structure within QCD that follow from
all these investigations.

\section{Goldstone boson loops and non-analyticity}
{}For our purposes the primary significance of spontaneous chiral
symmetry breaking in QCD is that there are  
contributions to hadron properties from loops involving the 
resulting Goldstone bosons.  These
loops have the unique property that they give rise to terms in an
expansion of most hadronic properties as a function of quark mass which
are not analytic.  As a simple example we consider the nucleon mass.  The
most important chiral corrections to $M_N$ come from the processes
$N \ra N\pi \ra N$ ($\sigma_{NN}$) and $N \ra \Delta \pi \ra N$
($\sigma_{N \Delta}$). (We will come to what it means to say these are
the most important shortly.)  We write
$M_N = M_N^{\rm bare} + \sigma_{NN} + \sigma_{N \Delta}$.
In the heavy baryon limit one has
\be
\sigma_{NN} = - \frac{3 g_A^2}{16 \pi^2 f_\pi^2}
\int_0^\infty dk \frac{k^4 u^2(k)}{k^2 + m_\pi^2},
\label{eq:3A}
\ee
where $g_A, f_\pi$ are strictly evaluated in the chiral limit.
Here $u(k)$ is a natural high momentum cut-off which is the Fourier
transform of the source of the pion field (e.g. in the cloudy bag model
(CBM) it is $3 j_1(kR)/kR$, with $R$ the bag radius \cite{CBM}). From
the point of view of PCAC it is natural to identify $u(k)$ with the
axial form factor of the 
nucleon, a dipole with mass parameter
$1.02 \pm 0.08$GeV.

Regardless of the form chosen for the ultra-violet cut-off, one
finds that $\sigma_{NN}$ is a non-analytic function of the quark mass.
The leading non-analytic (LNA) piece of $\sigma_{NN}$ is independent
of the form factor and gives
\be
\sigma_{NN}^{LNA} = - \frac{3 g_A^2}{32 \pi f_\pi^2} m_\pi^3
\sim \bar{m}^{\frac{3}{2}}.
\label{eq:4A}
\ee
This has a branch point, as a function of $\bar{m}$, at
$\bar{m} = 0$. Such terms can only arise from Goldstone boson loops.

\subsection{Case Study: the Nucleon Mass}
It is natural to ask how significant this non-analytic behaviour
is in practice.  If
the pion mass is given in GeV, $\sigma_{NN}^{LNA} = -5.6 m_\pi^3$
and at the physical pion mass it is just --17 MeV.
However, at only three times the physical pion mass, $m_\pi = 420$MeV,
it is --460MeV -- half the mass of the nucleon.  If one's aim is to extract
physical nucleon properties from lattice QCD calculations this is
extremely important.  The most sophisticated lattice calculations with
dynamical fermions are only just becoming feasible at such low masses
and to connect to the physical world one must extrapolate from
$m_\pi \sim 500$MeV to $m_\pi = 140$MeV.
Clearly one must have control of the chiral behaviour.

\begin{figure}[tbh]
\centering{\
\rotate{\epsfig{file=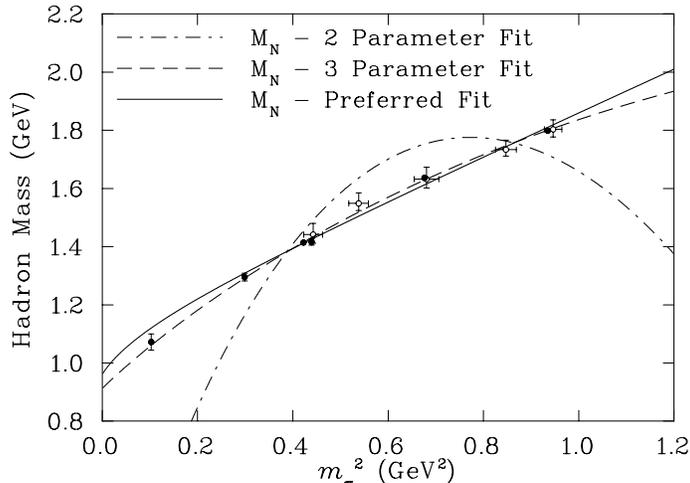,height=9cm}}
\caption{A comparison between phenomenological fitting functions for
the mass of the nucleon -- from Ref. \protect\cite{LEIN}.
The two parameter fit corresponds to using
Eq.(\ref{eq:5}) with $\gamma$ set equal to
the value known from $\chi$PT.  The three
parameter fit corresponds to letting $\gamma $
vary as an unconstrained fit parameter. The solid line is the
two parameter fit based on
the functional form of Eq.(\ref{eq:6}).
\label{fig:FIG2}}}
\end{figure}

{}Figure \ref{fig:FIG2} shows recent
lattice calculations of $M_N$ as a function of
$m_\pi^2$ from CP-PACS and UKQCD \cite{latt}.
The dashed line indicates a fit which
naively respects the presence of a LNA term,
\be
M_N = \alpha + \beta m_\pi^2 + \gamma m_\pi^3,
\label{eq:5}
\ee
with $\alpha, \beta$ and $\gamma$  fitted to
the data.  While this gives a very good fit to the data, the chiral
coefficient $\gamma$ is only -0.761, compared with the value -5.60
required by chiral
symmetry.  If one insists that $\gamma$ be consistent with QCD the best
fit
one can obtain with this form is the dash-dot curve.  This is clearly
unacceptable.

An alternative suggested recently by
Leinweber et al. \cite{LEIN}, which also
involves just three parameters, is to evaluate $\sigma_{NN}$ and
$\sigma_{N\Delta}$ with the same
ultra-violet form factor, with mass parameter $\Lambda$, and to fit
$M_N$ as
\be
M_N = \alpha + \beta m_\pi^2 + \sigma_{NN}(m_\pi,\Lambda) +
\sigma_{N\Delta}(m_\pi,\Lambda).
\label{eq:6}
\ee
Using a sharp cut-off ($u(k) = \theta(\Lambda - k)$) these
authors were able to obtain
analytic expressions for $\sigma_{NN}$ and $\sigma_{N\Delta}$
which reveal the correct LNA
behaviour -- and next to leading (NLNA) in the $\Delta \pi$ case,
$\sigma_{N\Delta}^{NLNA} \sim
m_\pi^4 \ln m_\pi$.
These expressions also reveal a branch point at $m_\pi = M_\Delta -
M_N$,
which is important if one is extrapolating from large values of $m_\pi$
to the physical value.  The solid curve in Fig.~1 is a two parameter fit
to the lattice data using Eq.(\ref{eq:6}), but fixing $\Lambda$
at a value suggested by CBM simulations to be
equivalent to the preferred 1 GeV dipole. A small increase in $\Lambda$
is necessary to fit the lowest mass data point, at $m_\pi^2 \sim
0.1$ GeV$^2$, but clearly one can describe the data very well while
preserving the exact LNA and NLNA behaviour of QCD.

\subsection{Consequences for the Sigma Commutator}
The analysis of the lattice data for $M_N$, incorporating the correct
non-analytic behaviour, can yield interesting new information concerning
the sigma commutator of the nucleon:
\be
\sigma_N = \frac{1}{3} \langle N| [Q_{i 5},[Q_{i 5},H_{QCD}]] |N\rangle
= \langle N| \bar{m} (\bar{u} u + \bar{d} d) |N\rangle.
\label{eq:8}
\ee
This is a direct measure of chiral SU(2) symmetry breaking in QCD, and
the widely accepted experimental value is
$45 \pm 8$MeV \cite{SIG_EX}. (Although there
are recent suggestions that it might be as much as 20 MeV
larger \cite{Kneckt}.)
Using the Feynman-Hellmann theorem one can also write
\be
\sigma_N = \bar{m} \frac{\partial M_N}{\partial \bar{m}}
= m_\pi^2 \frac{\partial M_N}{\partial m_{\pi}^2}.
\label{eq:9}
\ee
Historically, lattice calculations have evaluated
$<N| (\bar{u} u + \bar{d} d) |N>$ at large quark mass
and extrapolated this
scale dependent quantity to the ``physical'' quark mass, which had to
be determined in a separate calculation.  The latest result with
dynamical fermions, $\sigma_N = 18 \pm 5$ MeV \cite{SESAM},
illustrates how difficult this procedure is. On the other hand, if one
has a fit to $M_N$ as a function of $m_\pi$ which is
consistent with chiral symmetry, one can evaluate $\sigma_N$
directly using Eq.(\ref{eq:9}). Using Eq.(\ref{eq:6}) with a sharp
cut-off yields $\sigma_N \sim 55$ MeV, while a dipole form gives
$\sigma_N \sim 45$ MeV \cite{SIGMA}. The residual model dependence can
only be
removed by more accurate lattice data at low $m_\pi^2$. Nevertheless,
the result $\sigma_N \in (45,55)$ MeV is in very good agreement with the
data.  In
contrast, the simple cubic fit, with $\gamma$ inconsistent with chiral
constraints, gives $ \sim 30$ MeV. Until the experimental situation
regarding $\sigma_N$ improves, it is not possible to draw definite
conclusions regarding the strangeness content of the
nucleon. However,  
the fact that two-flavour QCD reproduces the current
prefered value should certainly stimulate some thought and a lot of work.

\section{Electromagnetic Form Factors}
\begin{figure}[t]
\centering{\
\epsfig{file=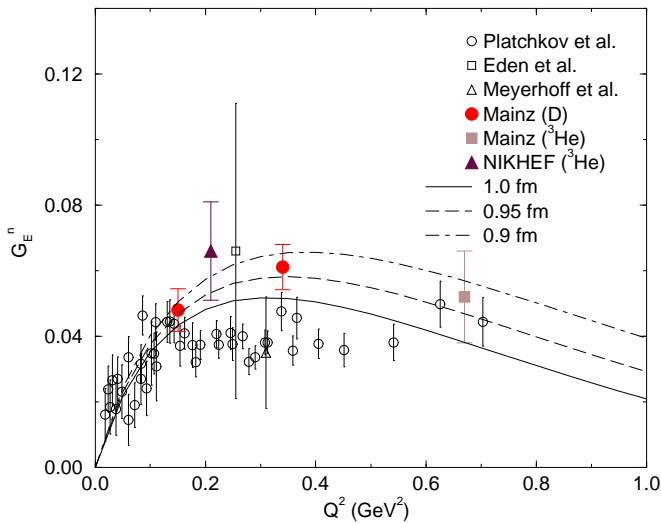,height=8cm}
\caption{Recent data for the neutron electric form factor
in comparison with CBM calculations for a
confining radius around 0.95fm -- from Ref.\ \protect\cite{LU}.}
\label{fig:neutron}}
\end{figure}
It is a general consequence of quantum mechanics that the
long-range charge structure of the proton comes from its $\pi^+$ cloud
($p \ra n \pi^+$),
while for the neutron it comes from its $\pi^-$ cloud ($n \ra p \pi^-$).
However, it is not
often realized that the LNA contribution to the nucleon charge radius
goes like $\ln m_\pi$ and diverges as $\bar{m} \ra 0$ \cite{LN}.
This cannot be reproduced by a
constituent quark model. Figure \ref{fig:neutron}
shows the latest data from Mainz and NIKHEF
for the neutron electric form factor, in comparison with CBM
calculations for a confinement radius between 0.9 and 1.0 fm. The
long-range $\pi^-$ tail of the neutron plays a crucial role.

\begin{figure}[ht]
\begin{center}
{\epsfig{file=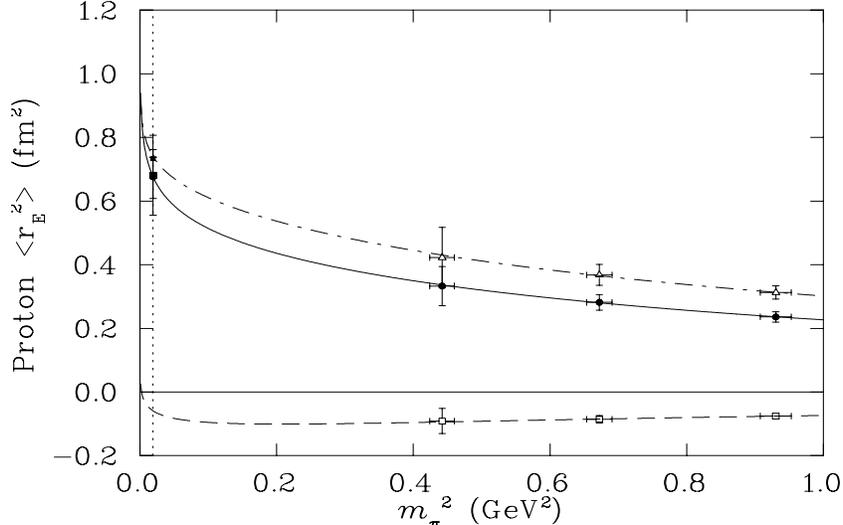, height=11cm, width=7cm, angle=90}}
\caption{Fits to lattice results for the squared electric charge radius
of the proton -- from Ref. \protect\cite{Emily}. Fits 
to the contributions from individual quark flavours are also
shown: the $u$-quark sector results are indicated by open triangles
and the $d$-quark sector results by open squares. Physical values
predicted by the fits are indicated at the physical pion mass, where
the full circle denotes the result predicted from the first
extrapolation procedure and the full square denotes the baryon radius
reconstructed from the individual quark flavor extrapolations. (N.B.
The latter values are actually so close as to be
indistinguishable on the graph.) The
experimental value is denoted by an asterisk.}
\label{fig:prot}
\end{center}
\end{figure}

While there are only limited (and indeed quite old) lattice data for
hadron charge radii, recent experimental progress in the determination
of hyperon charge radii has led us to examine the extrapolation
procedure for obtaining charge data from the lattice simulations
\cite{Emily}. Figure \ref{fig:prot} shows the extrapolation of the
lattice data \cite{LWD} for the charge radius of the proton. Clearly the
agreement with experiment is much better once the chiral log required by
chiral symmetry is correctly included, than if, for example, one simply
made a linear extrapolation in the quark mass (or $m_\pi^2$).
{}Full details of the results for all the octet baryons may be found in
Ref. \cite{Emily}.

The situation for baryon magnetic moments is also very interesting.
The LNA contribution in this case arises from the diagram where
the photon couples to the pion loop.  As this
involves two pion propagators the expansion of the proton and neutron
moments is:
\be
\mu^{p(n)} = \mu^{p(n)}_0 \mp \alpha m_\pi + {\cal O}(m_\pi^2).
\label{eq:10}
\ee
Here $\mu^{p(n)}_0$ is the value in the chiral limit and the
linear term in $m_\pi$ is proportional to $\bar{m}^{\frac{1}{2}}$,
a branch point at $\bar{m} = 0$.  The
coefficient of the LNA term is $\alpha = 4.4 \mu_N $GeV$^{-1}$.
At the physical pion mass this LNA
contribution is $0.6\mu_N$, which is almost a third of the neutron
magnetic
moment. {\em No constituent quark model can or should get better
agreement
with data than this.}

Just as for $M_N$, the chiral behaviour of $\mu^{p(n)}$ is vital to a
correct
extrapolation of lattice data. One can obtain a very satisfactory fit to
some rather old data, which happens to be the best available,
using the simple Pad\'e \cite{MAGMOM}:
\be
\mu^{p(n)} = \frac{\mu^{p(n)}_0}{1 \pm \frac{\alpha}{\mu^{p(n)}_0} m_\pi
+
\beta m_\pi^2}
\label{eq:11}
\ee
\begin{figure}[htb]
\begin{center}
{\epsfig{file=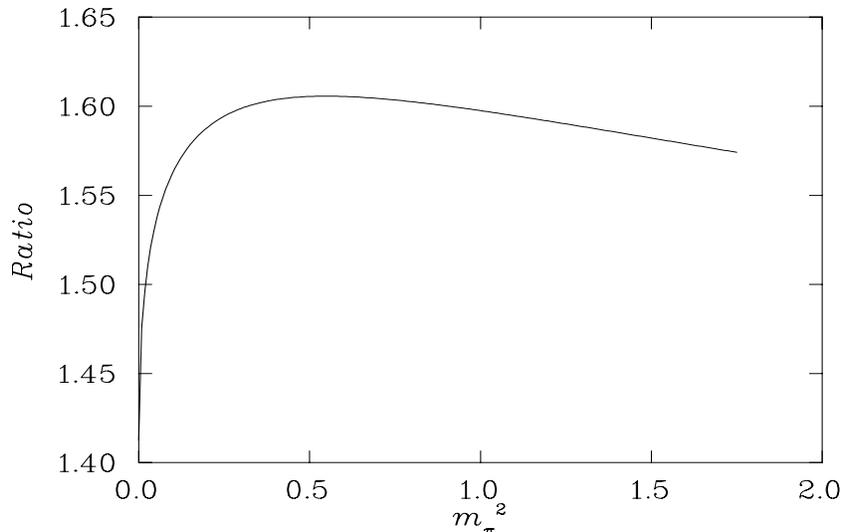, height=11cm, width=7cm, angle=90}}
\caption{Absolute value of the ratio of the proton 
to neutron magnetic moments as a function
of $m_\pi^2$ obtained from the Pad\'e approximants in Eq.
(\protect\ref{eq:11}). We stress that the behaviour as $m_\pi^2
\rightarrow 0$ is {\em model independent}.
}
\label{fig:ratio}
\end{center}
\end{figure}
The data can only determine two parameters and Eq.(\ref{eq:11})
has just two free parameters while guaranteeing the correct LNA
behaviour as $m_\pi \ra 0$ {\bf and} the correct behaviour of 
heavy quark effective theory (HQET)
at large $m_\pi^2$.  The
extrapolated values of $\mu^p$ and $\mu^n$
at the physical pion mass, 
$2.85 \pm 0.22 \mu_N$ and $-1.90 \pm 0.15 \mu_N$, respectively, 
are currently the best estimates from non-perturbative QCD
\cite{MAGMOM}. For more details of this fit we refer to Ref.
\cite{MAGMOM}, while for the application
of similar ideas to other members of the
nucleon octet we refer to Ref. \cite{HYPMM}, and for the
strangeness magnetic moment of the
nucleon we refer to Ref. \cite{StrMM}.

Incidentally, from the point of view of the naive quark model it is
interesting to plot the ratio of the absolute values of the 
proton and neutron magnetic moments
as a function of $m_\pi^2$. The agreement of the constituent quark
result, namely 3/2, with the experimental value to within a few percent 
is usually taken as a major success. However, we see from Fig.
\ref{fig:ratio} that it is in fact fortunate to obtain such close
agreement \cite{Ross}. We stress that the
large slope of the ratio near $m_\pi^2 = 0$ is {\em model independent}.

\section{Structure Functions}
The parton distribution functions (PDFs) of the nucleon are
light-cone correlation functions which, in the infinite
momentum frame, are interpreted as probability distributions for finding
specific partons (quarks, antiquarks, gluons) in the nucleon.
They have been measured in a variety of high energy processes, ranging
from deep-inelastic lepton scattering to Drell-Yan and massive vector
boson production in hadron--hadron collisions.
A wealth of experimental information now exists on spin-averaged PDFs,
and an increasing amount of data is being accumulated on spin-dependent
PDFs \cite{DATAREVIEW}.

At high momentum transfer ($Q^2$) the dominant component of the PDFs
are determined by non-perturbative matrix
elements of certain ``leading twist'' operators. In principle these
matrix elements, which correspond to moments of the measured structure
functions, contain vital information about the non-perturbative
structure of the target. 
An extensive phenomenology has been developed over the years within
model
QCD studies, and in some cases remarkable predictions have been made
from
the insight gained into the non-perturbative structure of the nucleon.
An example is the $\bar d-\bar u$ asymmetry, predicted \cite{AWT83} on
the basis of the nucleon's pion cloud \cite{EARLY}, which has been
spectacularly confirmed in recent experiments at CERN and Fermilab
\cite{EXPT}.
Other predictions, such as asymmetries between strange and antistrange
\cite{STRANGE} and spin-dependent sea quark distributions,
$\Delta \bar u - \Delta \bar d$, still await experimental confirmation.
Note that none of these could be anticipated without insight into the
non-perturbative structure of QCD.

Despite the phenomenological successes in correlating deep-inelastic and
other high energy data with low energy hadron structure, the {\em ad
hoc} nature of some of the assumptions made in deriving the low energy models
from QCD leaves open a number of questions about the ability to reliably
assign systematic errors to the model predictions.
One approach in which structure functions can be calculated
systematically
from first principles, and which at the same time allows one to search
{}for and identify the relevant low energy QCD degrees of freedom,
is lattice QCD.

\begin{figure}[htb]
\begin{center}
\rotate[l]{\epsfig{file=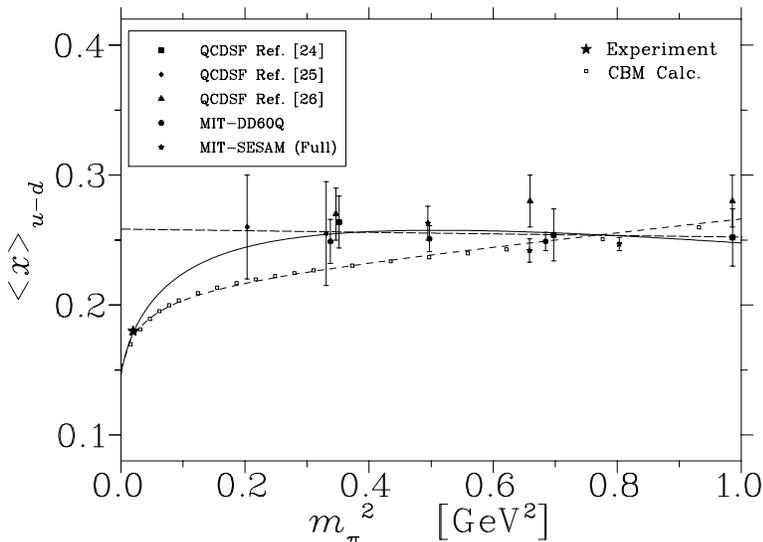,height=10cm}}
\caption{First moment of the difference $u-d$ from various lattice QCD
simulations (QCDSF [24--26] and MIT [27]), at a scale 
$Q^2=4$~GeV$^2$. Calculations from the
CBM are shown as small squares. 
The dashed curve is a simple fit which is linear in $m_\pi^2$,
while the solid curve incorporates the constraints of chiral symmetry,
as in Eq.(10).} 
\label{latt_mom}
\end{center}
\end{figure}
Early calculations of structure function moments within lattice QCD
were performed by Martinelli and Sachrajda \cite{MS}.
However, the most comprehensive analysis has been performed by the QCDSF
Collaboration [24--26] -- albeit within quenched QCD.
Recently the MIT group has performed the first full (unquenched) QCD
calculations of non-singlet moments \cite{DOLGOV}.
The moments from the full QCD simulations are very similar
to those from the quenched calculations. This is consistent with the
suggestions of chiral quark models, like the CBM, that in the mass
region currently accessible quark loops are suppressed.

As for the other nucleon properties discussed above, we propose to
extraplate the lattice data to the physical pion mass using a formula
which is compatible with the LNA structure of the PDFs. This behaviour
was derived recently, with the result that the LNA behaviour involved a
term in $m_\pi^2 \ln m_\pi$ \cite{TMS00}. For an initial investigation
we concentrate on the non-singlet combination of PDFs, $u - d$, in which
``disconnected'' quark loops cancel. Calculations based on the CBM
(which incorporate the LNA chiral structure just discussed) 
actually produce quite a reasonable description of the behaviour of the
moments of the PDFs as a function of quark mass, as shown in Fig.
\ref{latt_mom} (open squares). More important from the phenomenological
point of view, the CBM calculations (for the $n$'th moment of the PDFs)
can be fit with the simple expansion
in $m_\pi$:
\begin{equation}
\langle x^n_u - x^n_d \rangle = a_n + b_n m_\pi^2 + a_nc_{\rm LNA}
m_\pi^2
\ln \left( \frac{m_\pi^2}{m_\pi^2 + \mu^2} \right)\ ,
\label{eq:cmom}
\end{equation}
where $c_{\rm LNA}$ is model independent.
 
The scale $\mu$ in Eq.(\ref{eq:cmom}) is effectively the scale at which
the rapid, chiral variation at low $m_\pi$ turns off. The best fit to
the lattice data is obtained with a value $\mu \sim 0.4-0.5$ GeV -- a very
similar scale to that found, for example, for the magnetic moments.
Clearly Eq.(\ref{eq:cmom}) gives a very good description of the lattice
data for the first moment of the non-singlet distribution $d - u$. 
Taking into account the rapid chiral variation as $m_\pi^2 \rightarrow
0$ there is also quite good agreement between the extrapolated value
of the first moment and the experimentally determined moment. A similar
result holds for the second and third moments too \cite{TBP}.

\section{Conclusion}
In the light of the numerous examples presented in this brief review, it
should be evident that the study of hadron properties as a function of
quark mass shows a clear pattern:
\begin{itemize}
\item In the region of quark masses $\bar m > 60$ MeV or so ($m_\pi$
greater than typically 400-500 MeV)
hadron properties are smooth, slowly varying
functions of something like a constituent quark mass, $M \sim M_0 + c
\bar m$ (with $c \sim 1$).
\item Indeed, $M_N \sim 3 M, M_{\rho, \omega} \sim 
2 M$ and magnetic moments behave like $1/M$.
\item As $\bar m$ decreases below 60 MeV or so, chiral symmetry leads to
rapid, non-analytic variation, with $\delta M_N \sim {\bar m}^{3/2},
\delta \mu_H \sim {\bar m}^{1/2}$ and \\ $\delta <r^2>_{\rm ch} \sim \ln
\bar m$.
\item Chiral quark models like the cloudy bag provide a natural
explanation of
this transition. The scale is basically set by the inverse size of the
pion source -- the inverse of the bag radius in the bag model.
\end{itemize}

These are remarkable results that will have profound consequences for
our further exploration of hadron structure within QCD as well as the
analysis of the vast amount of data now being taken concerning unstable
resonances. In terms of immediate results for the structure of the
nucleon, we note that the careful incorporation of the correct chiral
behaviour of QCD into the extrapolation of its properties calculated on
the lattice has produced:
\begin{itemize}
\item The best values of the proton and neutron magnetic moments from
QCD.
\item The best value of the sigma commutator.
\item Improved values for the charge radii of the baryon octet.
\item Improved values for the magnetic moments of the hyperons.
\item Good agreement between the extrapolated moments of the non-singlet
distribution $u - d$ and the experimentally measured moments.
\end{itemize}
In addition, although we did not have time to discuss it, this approach
has led to the best current value for the strangeness magnetic moment of
the proton from lattice QCD \cite{StrMM}.

Clearly, while much has been achieved, even more remains to be done. It
is vital that lattice calculations with dynamical fermions are
pushed to the lowest possible quark masses, taking advantage of
developments of improved actions and so on. It is also vital to further
develop our understanding of the physics of chiral extrapolation by
comparison with these new calculations, by looking at new applications
and by further comparison with chiral models.

\begin{center}
Acknowledgements
\end{center}

We would like to thank E. Hackett-Jones, 
J. Negele, K. Tsushima, A. Williams and R. Young
for helpful discussions of the matters discussed here.
This work was supported by the Australian Research Council and the
University of Adelaide.

\end{document}